\documentclass{article} \usepackage{epsf} \usepackage{amsmath}
\usepackage{bbm}

\newcommand{\be}{\begin{equation}} 
\newcommand{\ee}{\end{equation}}

\newcommand{\D}{\mathcal{D}}

\newcommand{\psibar}{\overline{\psi}}

\newcommand{\chibar}{\overline{\chi}}

\newcommand{\I}{\mathrm{i}}

\newcommand{\com}[2]{[#1,#2]}
\newcommand{\anticom}[2]{\{#1,#2\}}

\begin{document}
\thispagestyle{empty} \parskip=12pt \raggedbottom

\noindent
\vspace*{1cm}
\begin{center}
  {\LARGE Lattice Regularization and Symmetries }

 
  \vspace{1cm} 
  Peter Hasenfratz, Ferenc Niedermayer${}^*$ and Reto von Allmen \\
  Institute for Theoretical Physics \\
  University of Bern \\
  Sidlerstrasse 5, CH-3012 Bern, Switzerland
  
  \vspace{0.5cm}
  
  \nopagebreak[4]
 
\begin{abstract}
Finding the relation between the symmetry transformations in the continuum 
and on the lattice might be a nontrivial task as illustrated by the history
of chiral symmetry. 
Lattice actions induced by a renormalization group procedure inherit all 
symmetries of the continuum theory. 
We give a general procedure which gives the corresponding symmetry 
transformations on the lattice.
\end{abstract}

\end{center}

\vfill
\noindent{---------------}\\
\noindent{\footnotesize ${}^*$On leave from E\"otv\"os University, 
HAS Research Group, Budapest, Hungary}

\eject

\section{Introduction}
In 1981 Ginsparg and Wilson formulated a condition \cite{GW1982} to be 
satisfied by the lattice Dirac operator in order to have the physical 
consequences of chiral symmetry on the lattice. 
The derivation of this condition is based on renormalization group (RG) 
considerations, but the result is more general. 
Indeed, the GW relation is satisfied not only by the fixed point
\cite{PH}, but also by the overlap \cite{Neu} and the domain-wall operator
after dimensional reduction \cite{Ki}. The Dirac operators in the latter two
cases are not related to RG ideas.

The GW condition is a non-linear relation for the lattice Dirac
operator reflecting the fact that concerning chiral symmetry 
the physical content of the classical
lattice theory is the same as that in the continuum. It was observed only many
years later that an exact symmetry transformation exists on the lattice
as well \cite{Lu}.

Discretizing a field theory by repeated RG block transformation 
-- or equivalently, by 'blocking out of continuum' \cite{Wie} -- 
has the advantage  that all symmetries of the continuum theory will 
be inherited by the lattice action -- even those which are explicitly 
broken by the block transformation.
The symmetry transformations are, however, different from those in
the continuum. We present here a general technique and a streamlined
procedure to find the form of the symmetry transformations and the symmetry
conditions (like the GW relation). 

We have to emphasize that talking about a symmetry transformation on the 
lattice we mean a symmetry of the lattice {\it action}, 
i.e. the classical field theory. In the quantum theory this transformation 
enters as a change of variable in the path integral which might induce 
a non-trivial contribution to the Ward identity by the integration measure. 

Not all internal symmetries of the continuum theory can be kept
by the block transformation. 
Consider, as an example, the chiral symmetry. 
For a continuum action and a block transformation which both have
an explicit $\gamma_5$-invariance, the resulting lattice action will also
be $\gamma_5$-invariant. But due to the existence of the chiral anomaly,
this action cannot be an acceptable one -- it has to be either non-local
or has to describe extra unwanted degrees of freedom (doublers)\cite{NN}.
This is what happens in the limit when the coefficient of the natural 
block transformation goes to infinity. 
In this limit the blocking becomes chiral invariant,
but at the same time the corresponding lattice action ceases to 
be local \cite{Uwe}.

This paper is motivated by some unsolved theoretical problems in lattice
regularized chiral gauge theories. In spite of the great progress during the
last years \cite{chth} an important problem remains: the relative weight
between the different topological sectors is undefined. This situation might 
be related to different technical problems.
Although the chiral invariant {\it vector} theory has a controlled RG
background, the steps towards a chiral theory are not related to RG anymore. 
The projectors \cite{proj} are introduced by hand and, seemingly unavoidably,
they break CP and T symmetry \cite{CP}. 
Further, the fermion number anomaly\footnote{We mean here the global vector
anomaly of a chiral gauge theory free of gauge anomalies.}  enters in
an unusual 
way: the different topological sectors have different number of degrees 
of freedom on the lattice. 
These technical issues might be related to the problem mentioned above.
A different strategy would be to start with a fermion number violating
block transformation, which makes the relation  between the continuum
and lattice symmetries non-trivial.
The systematic approach discussed here might be a useful tool in this
and similar problems.   

\section{Free massless fermions}
Since fermions enter quadratically even in the presence of gauge fields,
most of the equations below remain valid in the presence of interactions
as well.

The block transformation is a Gaussian integral which is equivalent to a
formal minimization problem:
\be \label{min} 
  \chibar \D \chi = \min_{ \psibar,\psi} \left\{ \psibar D \psi +
      \left(\chibar - \psibar\omega^\dagger \right) \left( \chi -
        \omega \psi \right) \right\} 
\ee
where the fermion fields $\psi_x$ and $\chi_n$ live in the continuum and on
the lattice respectively, $\omega_{nx}$ is the blocking matrix,
$D_{xx'}=(\gamma^\mu \partial_\mu)_{xx'}$ and $\D_{nn'}$
are the continuum and lattice Dirac operators. For the blocking we take a
flat, non-overlapping averaging
  \begin{eqnarray}\label{omega}
  \omega_{nx} &=& \left\{ \begin{array}{l} 1 \quad \mathrm{if} \quad x
        \in \mathrm{block}\;n \,,\\
        0 \quad \mathrm{otherwise}\,. \end{array} \right.
  \end{eqnarray}
With this choice one has 
$\sum_x \omega_{nx}\omega^\dagger_{xn'} = \delta_{nn'}$,
i.e. $\omega \omega^\dagger = 1$.

The minimizing fields $\psi_0 = \psi_0(\chi)$ and $\psibar_0 =
\psibar_0(\chibar)$ in eq.~(\ref{min}) are given by
\begin{eqnarray}
  \label{minfield}
  \psi_0(\chi) &=& A^{-1}\omega^\dagger\chi, \nonumber \\
  \psibar_0(\chibar) &=& \chibar\omega A^{-1} 
\end{eqnarray}
where
\be \label{A}
  A = D + \omega^\dagger\omega .
\ee
Inserting eq.~(\ref{minfield}) into eq.~(\ref{min}) gives the lattice 
Dirac operator
\be \label{lattD}
  \D = 1 - \omega A^{-1}\omega^\dagger.
\ee
From the equations above it is easy to derive the following useful relations
which will be used repeatedly in this work
\be
\begin{array}{r@{\:=\:}l@{\quad}r@{\:=\:}l}
  \label{relations}
  \omega  \psi_0(\chi) & (1-\D)\chi \,, &
  \psibar_0(\chibar)\omega^\dagger & \chibar(1-\D) \,, \\
  D\psi_0(\chi) & \omega^\dagger \D \chi \,, & \psibar_0(\chibar) D &
  \chibar \D \omega \,.\\
\end{array}
\ee
The Ginsparg-Wilson relation can be obtained then from eq.~(\ref{lattD}) by
using $\anticom {D}{\gamma_5}=0$ and the relations 
above\footnote{{}Note that with our choice of the coefficients in 
eq.~\eqref{min} the factor 2 appears in the GW relation. 
This is, of course, just a convention.}:
\be \label{GW}
 \anticom{\D}{\gamma_5} = 2\D \gamma_5 \D \,.
\ee 

We formulate now a general statement on the form of infinitesimal symmetry
transformations on the lattice.

\noindent
{\it Statement}\\
Let $\delta\psi$ and $\delta\psibar$ be the change 
of the corresponding continuum fields under an infinitesimal 
symmetry transformation which leaves invariant the 
continuum action $\psibar D \psi$.

Define the infinitesimal change of the lattice fields by 
\begin{equation}
  \label{lattr}
  \delta\chi = \omega\delta\psi_0(\chi) \,, \quad
  \delta\chibar = \delta\psibar_0(\chibar)\omega^\dagger \,.
\end{equation}
Then the lattice action $\chibar \D \chi$ is invariant
under this infinitesimal symmetry transformations.

\noindent
{\it Proof}\\
One can use the explicit equations above to show the statement. 

Replace $\psi_0(\chi)$ by $\psi_0(\chi + \delta\chi) -\delta\psi_0(\chi)$ 
and $\psibar_0(\chibar)$ by 
$\psibar_0(\chibar + \delta\chibar) -\delta\psibar_0(\chibar)$ 
on the r.h.s of eq.~(\ref{min}),
where $\delta\psi_0(\chi)$ is an infinitesimal continuum symmetry
transformation of $\psi_0(\chi)$ and 
$\delta\chi$ is a not yet defined infinitesimal change of $\chi$. 
Since 
$\psi_0(\chi + \delta\chi) -\delta\psi_0(\chi) = \psi_0(\chi) + 
\mathrm{infinitesimally small}$ and
$\psi_0(\chi)$ is the minimum of the r.h.s. of eq.~(\ref{min}), 
the change of the r.h.s. is {\it quadratically} small:
\begin{eqnarray}\label{proof1}
  \lefteqn{ \chibar \D \chi = \psibar_0 (\chibar + \delta\chibar) D 
    \psi_0(\chi + \delta\chi) +} \nonumber\\
  && \left[ (\chibar - \left(\psibar_0 (\chibar + \delta\chibar) -
      \delta\psibar_0 (\chibar)\right) \omega^\dagger \right] \left[
    (\chi - \omega \left( \psi_0 (\chi + \delta\chi) - \delta\psi_0
      (\chi)\right) \right] + \nonumber\\ &&
  \:\mathrm{quadratically \: small} 
\end{eqnarray}
where, in the first term on the r.h.s. of eq.~(\ref{proof1}), we used that
$\delta\psi_0(\chi)$ 
is a symmetry transformation in the continuum. 

We identify now 
\begin{equation} \label{dchi}
\delta\chi = \omega\delta\psi_0(\chi) \,, \quad 
\delta\chibar = \delta\psibar_0(\chibar)\omega^\dagger \,,
\end{equation}
which leads to 
\begin{eqnarray}\label{proof2}
  \chibar \D \chi &=& \psibar_0 (\chibar + \delta\chibar) D \psi_0(\chi
  + \delta\chi) + \nonumber\\
  && + \left[ \chibar + \delta\chibar - \psibar_0
    (\chibar + \delta\chibar) \omega^\dagger \right] \left[ \chi +
    \delta\chi - \omega \psi_0 (\chi + \delta\chi)\right]
\end{eqnarray}
up to quadratically small corrections.
Comparing eq.~(\ref{min}) and eq.~(\ref{proof2}) we obtain
\be
 (\chibar + \delta\chibar) \D (\chi + \delta\chi) = \chibar \D \chi
\ee
i.e. eq.~\eqref{dchi} 
is a symmetry transformation of the lattice action $\chibar \D \chi$.

\section{Symmetry transformations of the lattice action, examples}

\noindent
{ \it U(1) axial transformation}\\
The standard infinitesimal axial rotation in the continuum reads
\be
\label{StAc}
\delta\psi_0(\chi) = \I \epsilon \gamma_5 \psi_0(\chi), \quad
\delta\psibar_0(\chibar) = \I \epsilon \psibar_0(\chibar) \gamma_5 \,.
\ee
The corresponding lattice transformation has the form
\begin{eqnarray}
  \label{StAl}
  \delta \chi &=& \I \epsilon \gamma_5 \omega \psi_0(\chi) = \I \epsilon
  \gamma_5 (1-\D) \chi \,, \nonumber\\   
  \delta \chibar &=& \I \epsilon \psibar_0(\chibar) \gamma_5
  \omega^\dagger = \I \epsilon \chibar (1-\D)\gamma_5 \,,
\end{eqnarray}
where we used eq.~(\ref{lattr}) and eq.~(\ref{relations}). These
transformations have the well known form found by L\"uscher \cite{Lu}. Notice,
however that the axial transformation in the continuum is not unique. The
following transformation, for example, also leaves the continuum action
invariant 
\be
\label{modAc}
\delta\psi_0(\chi) = \I \epsilon \gamma_5 (1-\alpha D)\psi_0(\chi),
\quad \delta\psibar_0(\chibar) = \I \epsilon \psibar_0(\chibar) (1 +
\alpha D) \gamma_5 \,.
\ee
The associated lattice transformation reads
\begin{eqnarray}
  \label{modAl}
  \delta \chi &=& \I \epsilon \gamma_5 \left( 1 - (1 + \alpha)\D \right)
  \chi \nonumber\\   
  \delta \chibar &=& \I \epsilon \chibar \left( 1 - (1-\alpha)\D\right) 
  \gamma_5 \,.
\end{eqnarray}
The $\alpha=1$ case\footnote{{}as well as the $\alpha=-1$ case}
is special since $\gamma_5$ and 
$\hat{\gamma}_5 = \gamma_5 (1 - 2\D)$, (for which $\hat{\gamma}_5^2 = 1$) 
are candidates to build the lattice L/R projectors for $\chibar$ and $\chi$, 
respectively \cite{proj}. 
Notice the asymmetry between the transformations of $\chi$ and $\chibar$, 
which is the source of CP violation in the present formulation of
chiral gauge gauge theories mentioned in the introduction. 

\noindent
{ \it U(1) vector transformation}\\
The standard infinitesimal vector rotation in the continuum $\delta
\psi_0 (\chi) = \I \epsilon \psi_0 (\chi)$, $\quad  \delta
\psibar_0 (\chibar) = -\I \epsilon \psibar_0 (\chibar)$ implies the
lattice transformation
\be
\delta \chi = \I \epsilon (1 - \D) \chi, \quad \delta \chibar = -\I
\epsilon \chibar (1 - \D)
\ee
while the transformation $\delta \psi_0 (\chi) = \I \epsilon (1 -
\alpha D) \psi_0 (\chi)$, $\delta
\psibar_0 (\chibar) = -\I \epsilon \psibar_0 (\chibar) (1 - \alpha D)$ leads to
\begin{eqnarray}
  \delta \chi &=& \I \epsilon \left(1 - (1 + \alpha)\D\right) \chi
  \nonumber\\
  \delta \chibar &=& -\I \epsilon \chibar \left(1 - (1 + \alpha)\D\right).
\end{eqnarray}
For $\alpha=0$ the continuum, while for $\alpha=-1$ the lattice transformation
has the standard form.  

Considering finite transformations, note that for $\alpha=0$ 
in eqs.~\eqref{modAc}, \eqref{modAl} the transformation 
$\exp(it\gamma_5 (1-\D))$ is not compact
(it is not $2\pi$-periodic in $t$) as opposed to the corresponding 
transformation  $\exp(it\gamma_5)$  in the continuum.
On the other hand, for $\alpha=1$ (or $\alpha=-1$) the lattice 
transformation  $\exp(it\hat{\gamma}_5)$ corresponding to 
eq.~\eqref{modAl} is compact, while its continuum counterpart is not.

\noindent
{ \it Infinitesimal translation}\\
In the continuum we have $\delta \psi_0 (\chi) = \epsilon
\hat{\partial}_\mu \psi_0 (\chi)$, $\delta \psibar_0 (\chibar) = \epsilon 
\psibar_0 (\chibar)\hat{\partial}_\mu^\dagger$, 
where 
$\left( \hat{\partial}_\mu \right)_{xy} = \partial_\mu^x \delta (x-y)$. 
Our general procedure leads to the lattice transformations
\begin{eqnarray}
  \delta \chi &=& \epsilon \omega \hat{\partial}_\mu \psi_0 (\chi)
  \nonumber \\
  \delta \chibar &=& \epsilon \psibar_0 (\chibar) 
  \hat{\partial}_\mu^\dagger \omega^\dagger \,.
\end{eqnarray}
Using $\com{D}{\hat{\partial}_\mu} = 0$ 
it is a simple exercise to show explicitly that the lattice action is 
invariant under this infinitesimal translation. 

What was shown above remains valid also in the presence of gauge fields.
In the RG approach, eq.~\eqref{min}, the lattice Dirac operator $\D$
lives on some lattice gauge field background $V$, while on the 
r.h.s. the continuum Dirac operator $D$ and the blocking $\omega$
are defined on a corresponding continuum gauge field, which is
obtained from $V$ by a similar minimization procedure involving
only gauge fields \cite{DHHN}.

\section{Generalization to interactive theories: the non-linear
  $\sigma$ model} 
We illustrate the generalization of the technique used above on the example of
the $d=2$ nonlinear sigma model. The equation analogous to eq.~(\ref{min})
reads in this case 
\be
\label{mins}
\mathcal{A} (\vec{R}) = \min_{\{\vec{S}\}} \left( A(\vec{S}) + T
  \left( \vec{R},\: \omega (\vec{S}) \right) \right)
\ee
where $T$ is the block transformation, $\omega (\vec{S})$ defines 
the averaging, $ A(\vec{S})$ is the continuum action, 
while $\mathcal{A} (\vec{R})$ is the (fixed point \cite{FP}) action on the 
lattice. For the averaging one might take the flat, non-overlapping averaging
in eq.~(\ref{omega}). A simple example for $T$ is
\begin{eqnarray}
  \label{simple}
  T &=& 2\kappa \sum_n \left( \vec{R}_n - \frac{(\omega \vec{S})_n}
    {|(\omega \vec{S})_n|}\right)^2 \nonumber \\
    &=& 4\kappa \sum_n \left( 1 - \vec{R}_n \tilde{\omega} (\vec{S})_n \right)
\end{eqnarray}
where we introduced the notation
\be
\tilde{\omega} (\vec{S})_n = \frac{(\omega \vec{S})_n}
    {|(\omega \vec{S})_n|}, \quad \tilde{\omega} (\vec{S})^2 = 1.
\ee
For notational simplicity we shall take $4 \kappa =1$. The minimizing field in
eq.~(\ref{mins}) is denoted by $\vec{S}_0 = \vec{S}_0 (\vec{R})$. 
Consider now an infinitesimal symmetry
transformation of the continuum action (infinitesimal translation, for
example) acting on the minimizing field 
$\vec{S}_0 (\vec{R}) \rightarrow \vec{S}_0 (\vec{R}) + \delta \vec{S}_0
(\vec{R})$. Introduce the notation
\be
\tilde{\omega} (\vec{S}_0 + \delta \vec{S}_0) = \tilde{\omega}
(\vec{S}_0) + \delta \tilde{\omega}, \quad \left( \delta
  \tilde{\omega},\: \tilde{\omega}(\vec{S}_0) \right) = 0.
\ee
Following the procedure used for fermions above, we compensate the change of
the blocking term $T$ by changing the lattice configuration 
$\vec{R} \rightarrow \vec{R} + \delta \vec{R}$
\be
\label{cond}
\left( \delta \vec{R},\: \tilde{\omega}(\vec{S}_0 (\vec{R})) \right) +
\left( \vec{R},\: \delta \tilde{\omega} \right) = 0, \quad (\vec{R},\:
\delta \vec{R}) = 0
\ee
where we used eq.~(\ref{simple}). The solution of eq.~(\ref{cond}) can be
written as
\be
\label{solution}
\delta \vec{R} = \delta \tilde{\omega} \left( \tilde{\omega}, \vec{R}
\right) - \tilde{\omega} \left( \delta \tilde{\omega}, \vec{R} \right).
\ee
If $\delta \vec{S}_0 (\vec{R})$ is a symmetry transformation in the
continuum, then $\mathcal{A} (\vec{R} + \delta \vec{R}) = \mathcal{A}
(\vec{R})$,
i.e. $\delta \vec{R}$ in eq.~(\ref{solution}) is a symmetry transformation 
on the lattice.

{\bf Acknowledgements}
P.H. is indebted for the kind hospitality and for the discussions with
many of the participants at the ILFTN Workshop in
Nara and at the Workshop at YITP in Kyoto.
P.H. and F.N. thank the kind invitation to the 
Ringberg Meeting where part of this work was presented.
This work was supported by the Schweizerischer Nationalfonds.

\eject


\end{document}